\documentclass[twocolumn,showpacs,preprintnumbers,amsmath,amssymb,superscriptaddress,10pt,aps,prb]{revtex4-2}
\usepackage{epsfig,amsopn}
\usepackage{graphicx}
\usepackage{epstopdf}
\usepackage{tikz}
\usepackage[version=3]{mhchem}
\usepackage{float}
\usepackage{amsmath,amssymb}
\usepackage{amsthm}
\usepackage{pifont}
\usepackage{simplewick}
\usepackage{enumerate}
\usepackage{xcolor}
\usepackage[normalem]{ulem}
\usepackage{chemformula}
\usepackage{siunitx}
\usepackage{booktabs}
\newcommand{\xmark}{\ding{55}}%
\usepackage[colorlinks=true,linktoc=page,linkcolor=magenta,citecolor=magenta]{hyperref}

\def\checkmark{\tikz\fill[scale=0.4](0,.35) -- (.25,0) -- (1,.7) -- (.25,.15) -- cycle;}

\newcommand{\ket}[1]{\left| #1 \right>} 

\newcommand\bea{\begin{eqnarray}}
	\newcommand\eea{\end{eqnarray}}
\newcommand\beq{\begin{equation}}  
	\newcommand\eeq{\end{equation}}

\begin{document}
\title{Non-linear  magnon transport in a bilayer van der Waals antiferromagnets}	
\author{Rohit Mukherjee}
\email{rohitmk@iitk.ac.in}
\affiliation{Department of Physics, Indian Institute of Technology - Kanpur, Kanpur 208 016, India.}
\author{Sonu Verma}
\email{sonu.vermaiitk@gmail.com}
\affiliation{Center for Theoretical Physics of Complex Systems, Institute for Basic Science (IBS), Daejeon, 34126, Korea}
\author{Arijit Kundu}
\email{kundua@iitk.ac.in}
\affiliation{Department of Physics, Indian Institute of Technology - Kanpur, Kanpur 208 016, India.}

\begin{abstract}

In this paper, we study the Berry curvature induced linear and nonlinear magnon transport in bilayer van der Waals antiferromagnets, where we deduce forms for the spin and energy currents within the semiclassical Boltzmann formalism under the relaxation time approximation. Even in the absence of the  Dzyaloshinskii-Moriya interaction, if we turn on the layer-dependent electrostatic doping (ED) potential and anisotropy in the Heisenberg interactions, the linear response remains zero, whereas, we obtain a nonzero nonlinear thermal Hall response resulting from higher moments of the Berry curvature. We show that, there is a sign reversal of nonlinear thermal Hall conductivity with varying strength of ED potential, which can be potentially useful in spin-based technologies. We also comment on the  momentum and temperature dependence of the relaxation time which can influence the transport properties.
\end{abstract}

\maketitle

\section{Introduction}

Anomalous transport signatures as a consequence of the presence of Berry phase of electronic systems has been studied extensively in the past~\cite{RevModPhys.82.1959,berryxio1,berryxio2,berryxio3}. Berry phase driven non-vanishing transport signatures in the linear response regime requires broken time-reversal symmetry (TRS), thus, such anomalous transport have been under intense investigation, especially, in quantum Hall systems~\cite{PhysRevLett.45.494,PhysRevLett.49.405,doi:10.1126/science.1087128,RevModPhys.87.1213,PhysRevLett.95.226801,PhysRevLett.95.146802}. It is recently understood that even in time-reversal symmetric systems signatures of Berry curvature and other band geometric quantities can appear beyond the linear response. In particular, in a time-reversal symmetric but inversion broken system, due to the presence of Berry curvature dipole (BCD) in the reciprocal space, there can be non-trivial electrical as well as optical response in the nonlinear regime~\cite{PhysRevLett.115.216806,PhysRevLett.121.246403,PhysRevB.97.195151,PhysRevB.98.121109,PhysRevB.97.041101,PhysRevLett.123.246602}. Numerous studies have been carried out in the recent past of BCD related anomalous transport, which include nonlinear anamalous Hall~\cite{PhysRevLett.121.266601,PhysRevB.99.035403,PhysRevB.100.195117}, Nernst~\cite{PhysRevB.99.201410,PhysRevB.100.245102} and thermal Hall effects~\cite{PhysRevResearch.2.032066}.

In similarity to electronic systems, Berry curvature plays an important role in the transport properties of magnetic systems, where the transport is carried by quantized spin wave excitations or the magnons. In magnetic systems the presence of the Dzyaloshinskii-Moriya interactions (DMI) among the spins can generate complex hopping elements in the effective magnon Hamiltonian that makes the magnon bands topological, and hence, one finds the linear response coefficients to be nonzero~\cite{Xioprl,alexyprl}. In the absence of DMI, Berry curvature related transport appear only in the nonlinear response regime, as in the case of electronic systems, where the responses are due to the higher moments of Berry curvature. There are a few recent studies that addresses this problem, especially, in spin Seebeck effect~\cite{PhysRevB.98.020401}, spin-Nernst effect~\cite{PhysRevResearch.4.013186} and optical responses~\cite{PhysRevB.98.134422,PhysRevB.100.224411}, but there exists no study of thermal Hall response in the nonlinear regime of spin-systems, as far our knowledge.

In this paper, we investigate linear as well as nonlinear responses of the magnons in presence a temperature gradient in the semiclassical Boltzmann transport framework, where we find that the nonlinear thermal hall response can also be attributed to the presence BCD. We apply our calculation in a bilayer van der Waals honeycomb antiferromagnet with anisotropic Heisenberg interactions under the presence of a layer-dependent electrostatic doping potential (ED). Antiferromagnetic honeycomb lattices are excellent platforms for exploring magnon transport properties as these systems support collinear ground states. Previous work on honeycomb lattice antiferromagnet~\ch{MnPS_{3}} showed the existence of linear spin-Nernst current in the presence of DMI interaction~\cite{Xioprl,alexyprl}. Both for single layer and bilayer honeycomb lattices, the linear thermal hall current remains zero due to a global time-reversal symmetry. Though recent neutron scattering experiments~\cite{PhysRevB.103.024424} suggest that \ch{MnPS_{3}} has effectively zero DMI. Thus, one of the possible explanations for the observed magnon-Nernst~\cite{nernst_exp} response can be explained by the Berry curvature dipole-induced nonlinear currents which was studied in a recent paper~\cite{PhysRevResearch.4.013186}, and other possible mechanisms include the magnon-magnon and magnon-phonon coupling~\cite{PhysRevX.12.041031}. 
In our work, with finite ED, even in the absence of DMI, we obtain an anisotropy-induced nonlinear magnon thermal hall response, while the total nonlinear magnon spin-Nernst current remains zero. Interestingly, we also find a sign reversal of the nonlinear Hall conductivity with increasing strength of ED, which can have potential applications in spintronics. 

In addition to nonlinear response, we also study the linear spin-Nernst response in the same system with DMI and Heisenberg interactions terms up to the third-order (i.e, keeping $J_1$, $J_2$ as well as $J_3$ hoppings).  We comment on the possible temperature and momentum dependence of the magnon scattering time, leading to finite lifetime of these modes and their effects in the nonlinear transport properties.

Our results show a direct control of the responses of these magnetic systems by means of electrical doping, an emerging area of research with potential for application in quantum devices~\cite{matsukura2015control,sun2019electric}. The recent advance in the field of van der Waals heterostructures have also opened new avenues for such electrical control of magnetism~\cite{he2010robust}. Application of electrostatic doping (ED) technique has already been used to tune the local moments in atomically thin bilayer systems such as \ch{CrBr_{3}}~\cite{ahn2006electrostatic} and \ch{CrI_{3}}~\cite{jiang2018controlling}.

This paper is organized as follows: in Sec.~\ref{formalism}, we present the expression for the linear and nonlinear magnon spin-Nernst and Hall coefficients by invoking semiclassical Boltzmann transport formalism. Next in Sec.~\ref{physical system}, we introduce the model spin Hamiltonian, where we study different transport coefficients.  In Sec.~\ref{numerical results}, we provide the details of the numerical simulations and discuss the results. We conclude with further discussions and summary of our work in Sec.~\ref{discussion}.

\begin{figure}[tbph]
	\centering
	\includegraphics[width=0.6\linewidth]{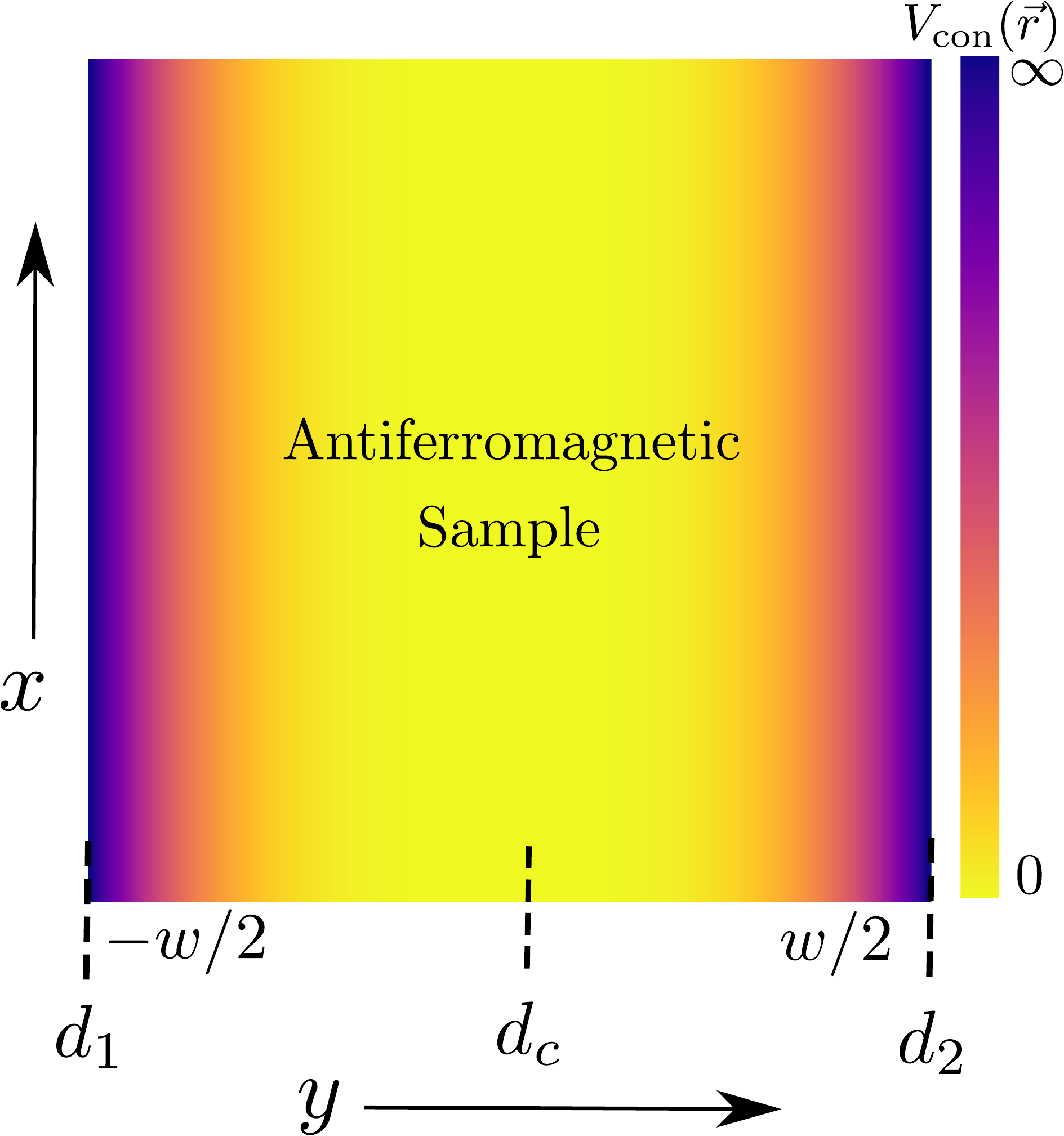}
	\caption{Setup used for the calculation of edge current in $x$ direction. $d_{c}$ and $d_{1},d_2$ are chosen well inside and outside the sample, respectively. The confining potential restricts the magnon wavepackets within the sample and its gradient exerts a force on the magnons which are described by Eq~(\ref{semiclass4}).}
	\label{fig:setup2}
\end{figure}

\section{Formalism}\label{formalism}
For the dynamics of quantum particles in a lattice, such as  electron (or magnon), we need the information of their dispersion as well as the Berry curvature of the Bloch bands~\cite{Xioprl}. The various transport properties can get considerably modified due to the presence of non-trivial Berry curvature (BC). The general properties of the Berry curvature of the band can be constrained by symmetry consideration. Under the time-reversal operation, the Berry curvature transform as $\Omega^{z}(\vec{k}) \to -\Omega^{z}(-\vec{k})$, on the other hand under the inversion  $\Omega^{z}(\vec{k}) \to \Omega^{z}(-\vec{k})$~\cite{PhysRevB.100.245102}. Thus, for a system with both the TRS and inversion symmetry, the Berry curvature vanishes identically over the whole Brillouin zone. The Chern number can be calculated by integrating the Berry curvature over the first Brillouin zone.




\begin{equation}
C_{n}=\dfrac{1}{2\pi}\int_{\rm BZ} d^2k \ \Omega_{n}^{z}(\vec{k}).
\end{equation}

We consider a magnon wave packet which is localized around the center, $r_{c},k_{c}$, in the real and the momentum space, respectively. The dynamics of the wavepacket is described by the semiclassical equations of motion (the suffix $c$ is omitted for brevity), which include an anomalous term due to the Berry curvature~\cite{PhysRevB.84.184406}:
\begin{equation}\label{semiclass3}
\dot{\vec{r}}=\dfrac{1}{\hbar}\dfrac{\partial E_n(\vec{k})}{\partial \vec{k}}- \dot{\vec{k}} \times \vec{\Omega}_{n}(\vec{k}),
\end{equation}
and,
\begin{equation}\label{semiclass4}
\hbar \dot{\vec{k}}= - \vec{\nabla} V_{\text{con}}(\vec{r}),
\end{equation}
here $n$ is the band index, $E_n(\vec{k})$ and  $\vec{\Omega}_{n}(\vec{k})$ are the energy and the Berry curvature of the $n^{th}$ magnon band in the momentum space, respectively.  The geometry we have is shown in Fig.~\ref{fig:setup2}, where we would like to find the current in the $x$-direction in response to a small temperature gradient in the $y$-direction. For the calculation of current, we follow the same line of derivation given in Ref.~\onlinecite{PhysRevB.84.184406}. The first term of the Eq.~(\ref{semiclass3}) describes the usual group velocity and the second term is the anomalous velocity arising from the Berry curvature of the magnon bands.  In electronic systems, the right-hand side of the Eq.~(\ref{semiclass4}) is usually the Lorentz force, but as the magnons are neutral quasiparticles, the force term can only be induced by a confining potential,  $\text{V}_{\rm con}(\vec{r})$, which we consider to be present only near the boundary of the antiferromagnetic sample. The confining potential restricts the magnon wavepacket within the sample and its gradient exerts the confining force. For the validity of Eq.~(\ref{semiclass3}) and Eq.~(\ref{semiclass4}), the spatial variation of the confining potential $V_{\text{con}}(\vec{r})$ should be much slower compared with the size of the magnon wave packet. If $w$ is the width of the sample, then we have
\begin{equation}
\begin{split}
&V_{\text{con}}(x,d_{c})=0 , \ \ V_{\text{con}}(x,d_{1})=V_{\text{con}}(x,d_{2})=\infty \\
&\text{with}, \ d_{1}<-w/2< d_{c}<w/2<d_{2},
\end{split}
\end{equation}
where $d_c$ is the center of the sample. 

\subsubsection*{Magnon current}
The averaged particle current density along the $x$ direction is given by,
\begin{equation}\label{eqsample}
\begin{aligned}
J_{x}&=\dfrac{1}{w} \int_{d_{1}}^{d_{2}} dy\   j_{x}(y)\\
&=\dfrac{1}{w} \int_{d_{c}}^{d_{2}} dy\   j_{x}(y)+\dfrac{1}{w} \int_{d_{1}}^{d_{c}} dy\   j_{x}(y).
\end{aligned}
\end{equation} 

Where $j_{x}(y)$ is the magnon current density in the the $x$ direction which is $y$ dependent. The confining potential varies slowly along the $y$ direction and $\dfrac{\partial V_{\text{con}}}{\partial y} \neq 0 $ only near $y=\pm w/2$. Thus,
\begin{equation}\label{boltz11}
\hbar \dot{\vec{k}}=-\dfrac{d V_{\text{con}}(y)}{dy} \hat{y}.
\end{equation}
The net velocity is then given by
\begin{equation}\label{boltz22}
\dot{\vec{r}}=\dfrac{1}{\hbar} \Big (\dfrac{\partial E_n(\vec{k})}{\partial k_{x}} \hat{x} +\dfrac{\partial E_n(\vec{k})}{\partial k_{y}} \hat{y} \Big) +\dfrac{1}{\hbar} \dfrac{dV_{\text{con}}(y)}{dy} \Omega^{z}_{n}(\vec{k}) \ (\hat{y} \times \hat{z}).
\end{equation}
The anomalous part of the velocity (second term) gives rise to magnon edge currents at the boundaries of the sample. 

The anomalous magnon current density in the $x$ direction is then given by,
\begin{equation}\label{magnoncurrent}
j^A_{x}(y)= \dfrac{1}{V} \sum_{n \vec{k}} \rho_{n} \big(\vec{k}, T(y)\big) \dfrac{1}{\hbar} \dfrac{dV_{\text{con}}(y)}{dy} \Omega^{z}_{n}(\vec{k}),
\end{equation}
where $\rho_n \big(\vec{k}, T(y)\big)$ is the non-equilibrium bosonic distribution function of the $n^{\rm th}$ band, $T(y)$ is the temperature as a function of the $y$ coordinate and $V$ is the area of the sample. Here we should mention that, apart from the velocity along the edge due to BC (second term in Eq.~(\ref{boltz22})) we have another contribution coming from the group velocity (first term in ~Eq.~(\ref{boltz22})) of the Bloch bands, so the magnon wavepackets may not move only along the edges. But, what we have written in Eq.~(\ref{magnoncurrent}) is indeed the total magnon edge current when all the magnons in the thermal equilibrium are added up, i.e, $j^A_x(y) \equiv j_x(y)$~\cite{PhysRevB.84.184406}.

Following the usual procedure, we write down the nonequilibrium distribution function as a sum of equilibrium distribution ($\rho^{(0)}$) and the first-order corrections due to temperature gradient. Details of the calculation are given in Appendix A,
\begin{widetext}
\begin{equation}\label{eq:currentd}
j_{x}(y)= \dfrac{1}{V}\sum_{n \vec{k}} \rho_{n}^{(0)}(E_n(\vec{k})+V_{\text{con}}(\vec{r});T(y))\dfrac{1}{\hbar} \dfrac{dV_{\text{con}}(y)}{dy} \Omega^{z}_{n}(\vec{k})+\dfrac{1}{V}\sum_{n \vec{k}} \rho_{n}^{(1)}(\vec{k};T(y))\dfrac{1}{\hbar} \dfrac{dV_{\text{con}}(y)}{dy} \Omega^{z}_{n}(\vec{k}).
\end{equation}
\end{widetext}
For the moment, we shall not discuss the first term of the above equation, which is the linear response of the system, instead, we shall focus on the second term, which is responsible for the nonlinear response:
\begin{equation}\label{densityx}
j^{\rm nl}_{x}(y)=\sum_{n \vec{k}} \dfrac{1}{V} \rho_{n}^{(1)}(\vec{k};T(y))\dfrac{1}{\hbar} \dfrac{dV_{\text{con}}(y)}{dy} \Omega^{z}_{n}(\vec{k}).
\end{equation}
Now we are in a position to calculate the non-equilibrium bosonic distribution function using the semiclassical Boltzmann transport equation under constant-relaxation-time ($\tau$) approximation~\cite{mahan2013many}, given as
\begin{equation}
\dot{\vec{r}}~\dfrac{\partial \rho}{\partial r}+\dot{\vec{k}}~\dfrac{\partial \rho}{\partial k}= -\dfrac{(\rho-\rho^{(0)})}{\tau}.
\end{equation}
Writing $\rho=\rho^{(0)}+\rho^{(1)}$ and after some straightforward algebra (given in Appendix A) we get the following form of the first order correction,
\begin{align}\label{rho1eq}
\rho_{n}^{(1)}=&\dfrac{-\tau}{\hbar}\Big(-\dfrac{E_n(\vec{k})-\mu}{T}\Big)\dfrac{\partial E_n(\vec{k})}{\partial k_{y}}
\dfrac{\partial \rho^{(0)}}{\partial E_n(\vec{k})}\dfrac{dT}{d y}\nonumber\\
&-\dfrac{\tau}{\hbar}\dfrac{\partial E_n(\vec{k})}{\partial k_{y}}\dfrac{\partial \rho^{(0)}}{\partial V_{\text{con}}}\dfrac{dV_{\text{con}}}{dy}+\dfrac{\tau}{\hbar}\dfrac{dV_{\text{con}}}{dy}\dfrac{\partial\rho^{(0)}}{\partial k_{y}}.
\end{align}
While calculating the current we neglect the contribution arising from the second and the third terms of Eq.~(\ref{rho1eq}), as they correspond to higher order corrections ($\mathcal{O}(\nabla T)^3$ and higher).

Now we plug the expression of Eq.~(\ref{rho1eq}) into Eq.~(\ref{eq:currentd}) to get the final form of the net magnon current density for the $n^{\rm th}$ Bloch band:
\begin{equation}
\begin{aligned}
j_{n,x}(y)&=\dfrac{1}{V} \sum_{\vec{k}}  \dfrac{1}{\hbar} \dfrac{dV_{\text{con}}(y)}{dy} \Omega^{z}_{n}(\vec{k})\rho^{(0)}_{n}\\
+&\dfrac{1}{V} \sum_{\vec{k}}  \dfrac{1}{\hbar} \dfrac{dV_{\text{con}}(y)}{dy} \Omega^{z}_{n}(\vec{k})\dfrac{\tau}{\hbar}\dfrac{E_n(\vec{k})-\mu}{T}\dfrac{\partial \rho^{(0)}_{n}}{\partial k_{y}} \nabla T,
\end{aligned}
\end{equation}
with $j_x(y) = \sum_n j_{n,x}(y)$, and $\nabla T \equiv\left(\dfrac{dT}{dy}\right)$.

Following further calculations (see Appendix A), we arrive at the following expression of the net averaged current density of the $n^{\rm th}$ band:
\begin{align}\label{finalform}
J_{n,x}=&\dfrac{k_{B}}{V} \sum_{\vec{k}}  \dfrac{1}{\hbar} \Omega^{z}_{n}(\vec{k})c_{1}(\rho^{(0)}_{n})(\nabla T)\nonumber \\
&+ \dfrac{1}{V} \sum_{\vec{k}}  \dfrac{1}{\hbar} \Omega^{z}_{n}(\vec{k})\dfrac{\tau}{\hbar}\dfrac{(E_n(\vec{k})-\mu)^{2}}{T^{2}}\dfrac{\partial \rho^{(0)}_{n}}{\partial k_{y}}(\nabla T)^{2},
\end{align}
with $J_x = \sum_n J_{n,x}$. Here $c_{\nu}$ are defined as
\begin{align}
c_{\nu}(\rho^{0}_{n}) =& -\int_{E_n(\vec{k})}^{\infty}(\epsilon\beta)^{\nu}  (\partial \rho^{(0)}_{n}/ \partial \epsilon)d\epsilon\nonumber\\
 =& \int_{0}^{\rho^{(0)}_{n}}\log\left[\left(\dfrac{1+t}{t}\right)\right]^{\nu}dt.	\label{eq:cnu}
\end{align}

The first and second terms in Eq.~(\ref{finalform}) correspond to the linear and nonlinear contributions of magnon current in the $x$ direction under the influence of a temperature gradient in the $y$ direction, respectively. We should note that the second term in Eq.~(\ref{finalform}) can be recast into the following form,
\begin{equation}
\dfrac{1}{V} (\nabla T)^{2} \sum_{\vec{k}}\dfrac{\tau}{\hbar^2 T}c_{1}(\rho^{0}_{n})\dfrac{\partial}{\partial k_{y}}[E_n(\vec{k})\Omega^{z}_{n}] ,
\end{equation} 
which also agrees with the result of Ref.~\onlinecite{PhysRevResearch.4.013186}. The quantity within the square bracket is termed an extended Berry curvature dipole which has similar implications to the BCD in electronic systems~\cite{PhysRevB.99.201410}.

\subsubsection*{Nernst, and thermal Hall current}
The magnon spin-Nernst current is defined as
\begin{equation}\label{nonlinearNersnt}
J_{x}^{\text{\rm Nernst}}=\hbar \sum_{n} \langle S^{z}_{n} \rangle J_{n,x},
\end{equation}
where $ \langle S^{z}_{n} \rangle$ is the expection value of $\hat{S}^{z}$ operator in $n^{\rm th}$ magnon band. Energy current for $n^{\rm th}$ band is simply given by (see Appendix A),
\begin{align}\label{nonlinearHall}
J^{\rm Energy}_{n,x}=&\dfrac{k_{B}^2T}{\hbar V} \sum_{\vec{k}}\Omega^{z}_{n}(\vec{k})c_{2}(\rho^{(0)}_{n})(\nabla T)\nonumber\\
&+ \dfrac{1}{V} \sum_{\vec{k}}   \Omega^{z}_{n}(\vec{k})\dfrac{\tau}{\hbar^{2}}\dfrac{(E_n(\vec{k})-\mu)^{3}}{T^{2}}\dfrac{\partial \rho^{(0)}_{n}}{\partial k_{y}}(\nabla T)^{2}.
\end{align}
The net magnon thermal Hall current is defined as the sum of the contribution arising from each band,
\begin{equation}\label{nonlinearHall2}
J_{x}^{\rm Hall}=\sum_{n} J^{\rm Energy}_{n,x}.
\end{equation}

It is instructive to note that these final expressions have similar forms as in the case of fermionic systems~\cite{nandi2}.

\subsubsection*{Nonvanishing transport coefficients based on symmetry}
Now, having arrived at this result, we present a short discussion on the symmetries of the dispersion and the Berry curvature, and their consequences on various terms in Eq.~(\ref{finalform}) and Eq.~(\ref{nonlinearHall}). For a time-reversal symmetric system, the Berry curvature is an odd function of $\vec{k}$ and the dispersion is an even function of $\vec{k}$ and hence the first terms of  both the Eq.~(\ref{finalform}) and  Eq.~(\ref{nonlinearHall}) are odd functions of $\vec{k}$, thus total contribution will vanish for each band when we sum over the entire BZ. But the second terms (which is the nonlinear contribution) for each of the equations are even functions under the exchange $\vec{k} \to -\vec{k}$ because of the presence of the term $\dfrac{\partial}{\partial k_{y}}$, and, as a consequence, the contribution coming from each band can be nonzero when we sum over the entire BZ. Overall spin-Nernst current and Hall currents, which are described by the equations (\ref{nonlinearNersnt}) and (\ref{nonlinearHall2}), respectively, can be non vanishing depending on the sign of $\langle S^{z}_{n} \rangle$ and the symmetry of Berry curvature. 
Below we take a generic spin Hamiltonian and analyze the above mentioned magnon transport coefficients.

\vspace{.5cm}
\section{van der Waals Honeycomb Antiferromagnet}\label{physical system}
\begin{figure*}[t]
	\centering
	\includegraphics[width=1.0\linewidth]{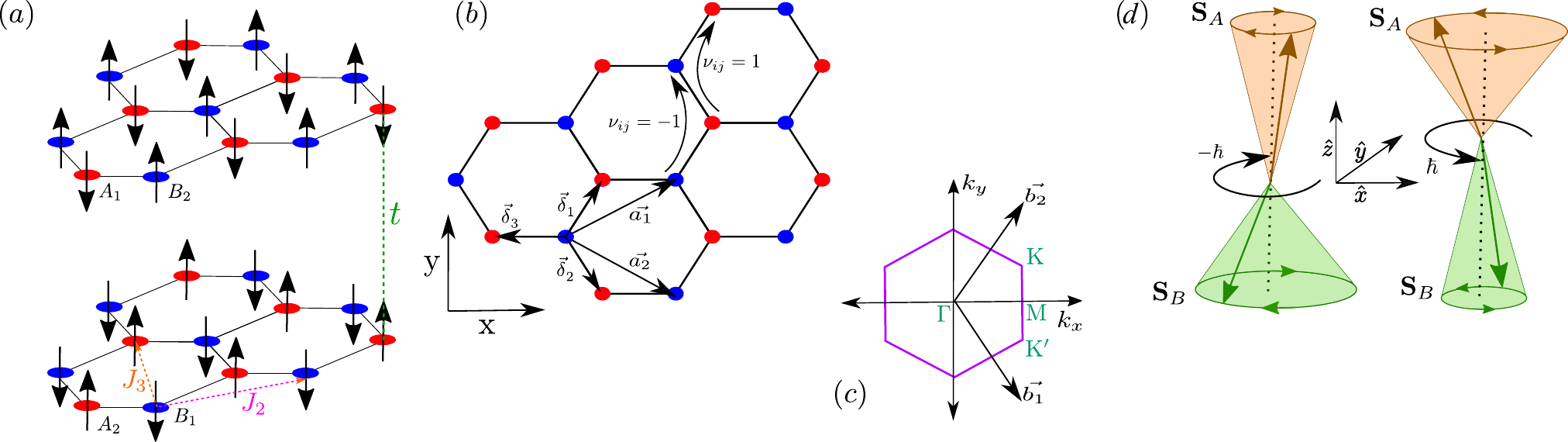}
	\caption{(a) Schematics of stacked bilayer honeycomb lattice. We are taking antiferromagnetic intra-layer Heisenberg interactions up to third order, in-plane easy axis anisotropy in both layers,  an antiferromagnetic inter-layer coupling and oppositely directed ED potentials. Red and blue circles correspond to the $A$ and $B$ sublattices respectively. (b) schematics of single layer honeycomb lattice in real space,  and, (c) unit cell in the reciprocal space. The real space and momentum space lattice vectors are also marked in these figures. (d) Schematic of the right and left-handed magnon modes in a single layer. The brown and green arrows represent the precession of spins on $A$/$B$ sublattices, in each mode the $S_{z}$ component is different for $A$ and $B$ sublattices during the spin wave precession, as a result, eigenmodes carry opposite spin-angular momentum (figure adapted from Ref.~\onlinecite{daniels2018nonabelian})}
	\label{fig:highsymmetry}
\end{figure*}

We take a stacked bilayer honeycomb lattice as our model Fig.~\ref{fig:highsymmetry}(a) to calculate the magnon transport coefficients. We consider a Hamiltonian of the spins consisting of various kinds of spin-spin interactions, which are relevant in van der Waals magnets~\cite{Ghader1,Xioprl}. The Hamiltonian is given by,
\begin{widetext}
\begin{equation}\label{eq:Hamiltonian}
\begin{aligned}
H=&\sum_{l=1,2}\left(\sum_{\langle i,j\rangle}J_{1j}\vec{S}_{i,l} \cdot \vec{S}_{j,l}+
\sum_{\langle \langle ij \rangle \rangle }J_{2}\vec{S}_{i,l} \cdot \vec{S}_{j,l}+
\sum_{\langle \langle \langle ij \rangle \rangle \rangle}J_{3}\vec{S}_{i,l} \cdot \vec{S}_{j,l}\right)\\
&+\sum_{
	\langle ij \rangle}t \ \vec{S}_{i,1} \cdot \vec{S}_{j,2}
+\sum_{l=1,2}\left(\sum_{i}K (S_{i,l}^{z})^{2}+ \sum_{i}U_{l}S^{z}_{i,l}+D\sum_{\langle \langle ij \rangle \rangle} \nu_{ij} [\vec{S}_{i,l} \times \vec{S}_{j,l}]_{z}\right),
\end{aligned}
\end{equation}
\end{widetext}
where $\vec{S}_{i,l}$ stands for the spin operator at site $i$ in the layer $l=1/2$. The first three terms within the braces consist of an antiferromagnetic Heisenberg interaction up to third order, the fourth term is an antiferromagnetic inter-layer coupling ($t$) between the nearest inter-layer sublattices, the fifth term is an easy-axis anisotropy term in each layer ($K$), the sixth is the layer dependent Electrostatic doping potential ($U_{l}$)~\cite{Ghader1} interaction, and the last term is the intra-layer DMI strength ($D$) among the second nearest-neighbors (the DMI coupling between the nearest neighbor spins vanishes as the inversion center of the Honeycomb lattice coincides with the center of the link joining the $A-B$ sublattice). Sign structure $\nu_{ij}$ is depicted in Fig~\ref{fig:highsymmetry}(b).
Further, we consider anisotropic nearest neighbor Heisenberg model where $J_{11}\neq J_{12} \neq J_{13}$, which might be induced by pressure in a realistic system~\cite{PhysRevResearch.4.013186}, whereas in absence of such anisotropy, $J_{11} = J_{12} = J_{13} \equiv J_1$. $\vec{\delta}_j$ ($j=1,2,3$) are the set of three vectors that connects the nearest neighbour sites. An easy axis anisotropy interaction $K$ stabilizes the N\'{e}el ordering in the $z$ direction.  $U_{l}=\pm U$ for layer $l=1/2$, respectively, which can be controlled by external doping of impurity ions.  

We proceed by writing down the Hamiltonian in terms of Holstein-Primakoff bosons defined as
\begin{widetext}
\begin{equation}
\text{Sublattice} \ A:\begin{cases}
\text{layer~1}: \ \ \hat{S}^{z}_{i,1}  \approx S-\hat{a}^{\dagger}_{i,1}\hat{a}_{i,1}, \  S_{A1}^{+} \approx \sqrt{2S}a_{1}, \ S_{A1}^{-}\approx \sqrt{2S}a^{\dagger}_{1}\\
\text{layer~2}: \ \ \hat{S}^{z}_{i,2} \approx -S+\hat{a}^{\dagger}_{i,2}\hat{a}_{i,2}, \  S_{A2}^{+} \approx \sqrt{2S}a^{\dagger}_{2}, \ S_{A2}^{-}\approx \sqrt{2S}a_{2}
\end{cases}       
\end{equation}
\begin{equation}
\text{Sublattice} \ B:\begin{cases}
\text{layer~1}: \ \ \hat{S}^{z}_{j,1} \approx -S+\hat{b}^{\dagger}_{j,1}\hat{b}_{j,1}, \  S_{B1}^{+} \approx \sqrt{2S}b^{\dagger}_{1}, \ S_{B1}^{-}\approx \sqrt{2S}b_{1}\\
\text{layer~2}: \ \ \hat{S}^{z}_{j,2} \approx S-\hat{b}^{\dagger}_{j,2}\hat{b}_{j,2}, \  S_{B2}^{+} \approx \sqrt{2S}b_{2}, \ S_{B2}^{-}\approx \sqrt{2S}b^{\dagger}_{2}
\end{cases}       
\end{equation}	
\end{widetext}

Fourier transformed operators are defined as,
\begin{equation}
\begin{bmatrix}
\hat{a}_{i} \\
\hat{b}_{i}
\end{bmatrix}=\dfrac{1}{\sqrt{N}}\sum_{k} e^{i \vec{k} \cdot \vec{r}}
\begin{bmatrix}
\hat{a}_{k} \\
\hat{b}_{k}
\end{bmatrix},
\end{equation}
where $N$ is the number of unit cells. Now the Hamiltonian can be written in the following form,
\begin{equation}
H=\dfrac{1}{2} \sum_{k} \Psi^{\dagger}(\vec{k})H(\vec{k})\Psi(\vec{k})
\end{equation}
where the full basis is given by,
\begin{equation*}
\Psi(\vec{k})=\Big[a_{1,\vec{k}} \ b_{1,\vec{k}} \ a^{\dagger}_{1,-\vec{k}} \ b^{\dagger}_{1,-\vec{k}} \ a_{2,\vec{k}} \ b_{2,\vec{k}} \ a^{\dagger}_{2,-\vec{k}} \ b^{\dagger}_{2,-\vec{k}}\Big]^{T},
\end{equation*}
where $a_{l} (b_{l})$ indicates the bosonic magnon annihilation operator at sublattice $A(B)$ in layer $l$ (details in Appendix B). 

\subsubsection*{Diagonalization and spectrum}
Our Hamiltonian in Eq.~(\ref{eq:Hamiltonian}) preserves the rotational symmetry along the $z$ direction (in the spin space). In this case, $[S^{z}_{\rm total},H]=0$, where $S^{z}_{\rm total} = \sum_{l,i}S^z_{i,l}$ is a good quantum number. We make a unitary transformation ($W$) of our basis such that the Hamiltonian becomes block diagonal with each block corresponding to a fixed $S^{z}$ sector. With
  \begin{equation*}
 \Psi^{\prime}(\vec{k})=W \Psi(\vec{k}),
 \end{equation*}
 our transformed hamiltonian becomes,
 \begin{equation}
 \begin{aligned}
 H=&\dfrac{1}{2}\sum_{k} \Psi^{\prime\dagger}(\vec{k})(U^{-1})^{\dagger}H(\vec{k})U^{-1}\Psi^{\prime}(\vec{k})\\
 &=\begin{bmatrix}
 H_{\uparrow} && 0\\
 0 && H_{\downarrow}
 \end{bmatrix},
 \end{aligned}
 \end{equation}
where,
\begin{align}
 &H_{\uparrow}=\dfrac{1}{2}\sum_{k} \Psi^{\prime \dagger}_{\uparrow}(\vec{k})H_{\uparrow}(\vec{k})\Psi^{\prime}_{\uparrow}(\vec{k}),\\
 &H_{\downarrow}=\dfrac{1}{2}\sum_{k} \Psi^{\prime \dagger}_{\downarrow}(\vec{k})H_{\downarrow}(\vec{k})\Psi^{\prime}_{\downarrow}(\vec{k}).
\end{align}
Here,
 \begin{widetext}
 \begin{equation}\label{mainH}
 H_{\uparrow}(\vec{k})=\begin{bmatrix}
 A+F-U+D &&0  && B-iC && t\\
 0 && A+F+U-D &&t && B+iC\\
 B+iC && t&& A+F+U-D&& 0\\
 t && B-iC && 0 &&A+F-U+D
 \end{bmatrix},
 \end{equation}
\end{widetext}
where $A=S(J_{11}+J_{12}+J_{13}+3J_{3}+t)$ , $\gamma_{\vec{k}}=S(J_{11}e^{ik_{x}/\sqrt{3}}+J_{12}e^{-i/2(k_{y}+k_{x}/ \sqrt{3})}+J_{13}e^{-i/2(-k_{y}+k_{x}/ \sqrt{3})})$, $g_{\vec{k}}=J_{3}S(e^{-2ik_{x}/ \sqrt{3}}+2e^{-ik_{x}/\sqrt{3}}\cos(k_{y}))$,$F=J_{2}S\Big[2(\cos k_{y}+\cos[-k_{y}/2-(\sqrt{3}/2)k_{x}]+\cos[-k_{y}/2+(\sqrt{3}/2)k_{x}])-6\Big]$,
$D=2D_{2}S[\sin(k_{y})+\sin(1/2(k_{y}+\sqrt{3}k_{x}))+\sin(1/2(k_{y}-\sqrt{3}k_{x}))]$.   
Where, $B=\text{Re} [\gamma_{\vec{k}}+g_{\vec{k}}]$ and $C=\text{Im} [\gamma_{\vec{k}}+g_{\vec{k}}]$. The basis for $\uparrow$ sector is given as,
 \begin{equation}
 \Psi^{'}_{\uparrow}(\vec{k})=(a_{\vec{k},1} \ b_{\vec{k},2} \ b^{\dagger}_{-\vec{k},1} \ a^{\dagger}_{-\vec{k},2} )^{T},
 \end{equation}
and, in a similar fashion, the  basis for the $\downarrow$ sector is given by,
 \begin{equation}
 \Psi^{'}_{\downarrow}(\vec{k})=(a_{\vec{k},2} \ b_{\vec{k},1} \ b^{\dagger}_{-\vec{k},2} \ a^{\dagger}_{-\vec{k},1} )^{T}.
 \end{equation}

 Now, in order to diagonalize the Hamiltonian in Eq.~(\ref{mainH}), we employ the standard technique of Bogoluivob transformation for quadratic bosonic Hamiltonian~\cite{bogo1,mukherjee2021schwinger}. We introduce new creation and annihilation magnon operators ($ \alpha^{\dagger} / \alpha$, \ $\beta^{\dagger}/ \beta $), such that,
\begin{equation}
\Psi^{\prime}_{\uparrow}(\vec{k})=\mathcal{T}_{\uparrow} \  \Gamma^{\prime}_{\uparrow}(\vec{k}); \ \Gamma^{\prime}_{\uparrow}(\vec{k})=(\alpha_{\vec{k},1} \ \beta_{\vec{k},2} \ \beta^{\dagger}_{-\vec{k},1} \ \alpha^{\dagger}_{-\vec{k},2} )^{T}. 
\end{equation}
We choose $\mathcal{T}_{\uparrow}$ such that the matrix $\mathcal{T}_{\uparrow}^{\dagger}H_{\uparrow}(\vec{k})\mathcal{T}_{\uparrow}$ becomes diagonal with the condition that $\mathcal{T}_{\uparrow} \ \Sigma_{z} \ \mathcal{T}_{\uparrow}^{\dagger}=\Sigma_{z}$ with $\Sigma_{z}=\sigma_{z} \otimes I_{2}$, where $\sigma_z$ is the Pauli matrix for the spin-space and $I_2$ is the identity in the layer-space. The last condition preserves the bosonic commutation rules in the new basis. The elements of the matrix $\mathcal{T}_{\uparrow}$ can be found from the eigenspectrum of the matrix $\Sigma_{z} H_{\uparrow}(\vec{k})$ which is also known as the dynamic matrix. More details of the procedure can be found in the Ref~\onlinecite{bogo2,mukherjee2021schwinger}. Similarly, we can diagonalize the Hamiltonian for the $\downarrow$ sector. After the diagonalization, we obtain four magnon bands, and corresponding eigen-kets $\ket{\alpha_{\vec{k},1}},\ket{\alpha_{\vec{k},2}},\ket{\beta_{\vec{k},1}},\ket{\beta_{\vec{k},2}}$.

\begin{equation}
\begin{aligned}
&\ket{\alpha_{\vec{k},l}}=\alpha_{\vec{k},l}^{\dagger}\ket{0}, 
\ket{\beta_{\vec{k},l}}=\beta_{\vec{k},l}^{\dagger}\ket{0},\\
&\alpha_{\vec{k},l} \ket{0}=0,
 \beta_{\vec{k},l} \ket{0}=0.
\end{aligned}
\end{equation}

\begin{figure*}[ht]
	\includegraphics[width=1.0\linewidth]{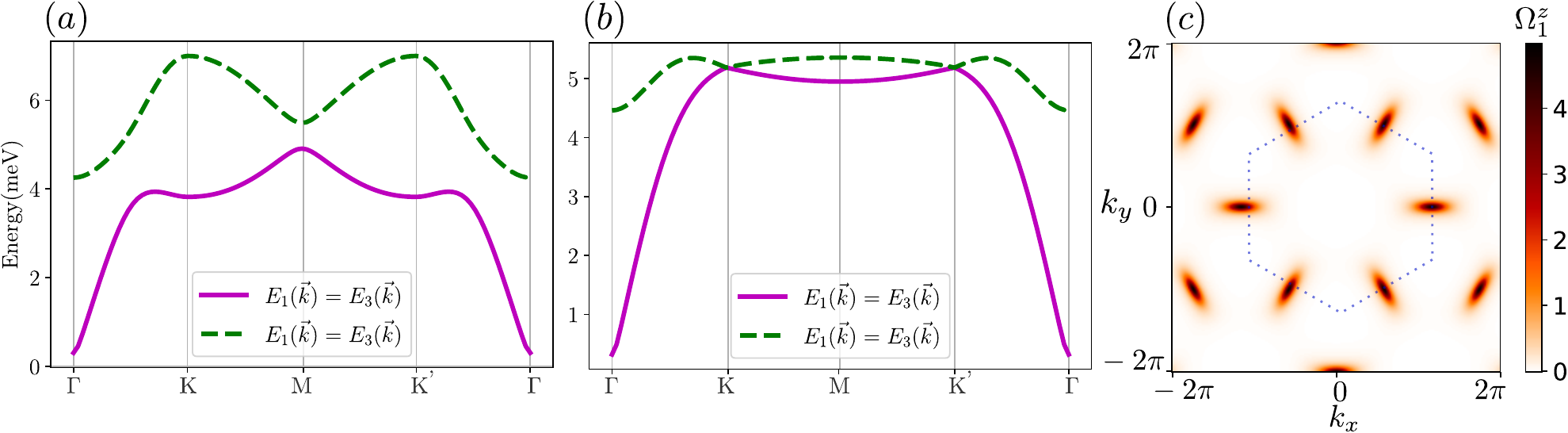}
	\caption{Model with $U=0$: In figures (a), (b), and (c) we plot the dispersion along the high symmetry points and the Berry curvature (lowest magnon band) for a model with Heisenberg interaction under the presence of the second nearest neighbor DM and easy-axis anisotropy. Parameters for the plot (a) and (c): $S$=1, $J_{1}=1.5 \  \text{meV}$, $J_{2}=0.05 J_{1}$, $J_{3}=0.1 J_{1}$, $K$=0.0086 \text{meV}, $U$=0, $t$=1.0 \text{meV}, $D$=0.3 \text{meV}. The values of the parameters are close to the real values in most of the van der Waals magnets~\cite{PhysRevB.101.205425} from the predictions by \textit{ab initio} calculations.  The dispersion is highly anisotropic as a function of $k$. With $D=0$ there is a band touching near $K$, $K'$ points (fig (b)). The density plot of the Berry curvature shows that it is highly concentrated near $M$ points and topological charges for each $M$ point is 1/3.  The dotted lines indicate the first Brillouin zone.}
	\label{fig:lineardnew1}
\end{figure*}

A schemetic diagram of the precesssion of spins in each layer for each magnon mode is depicted in Fig~\ref{fig:highsymmetry}(d). For either $D=0$ or $U=0$ the Hamiltonian for $H_{\uparrow}(\vec{k})$ (up-spin sector) and	$H_{\downarrow}(\vec{k})$ (down-spin sector) are related by, 
 \begin{equation*}
 H_{\uparrow}(\vec{k})=H_{\downarrow}^{*}(-\vec{k}).
 \end{equation*}
 The total spin-angular momentum can be written as,
\begin{equation}
S^{z}_{\rm total}= \sum_{i,l=1,2}S^{z}_{i,l,A} + S^{z}_{i,l,B},
\end{equation}
which we can write as,
\begin{equation}
\begin{aligned}
S^{z}_{\rm total}=&\sum_{i}(a^{\dagger}_{i,1}a_{i,1}-a^{\dagger}_{i,2}a_{i,2}-b^{\dagger}_{i,1}b_{i,1}+b^{\dagger}_{i,2}b_{i,2})\\
=&\sum_{k}(a^{\dagger}_{\vec{k},1}a_{\vec{k},1}-a^{\dagger}_{\vec{k},2}a_{\vec{k},2}-b^{\dagger}_{\vec{k},1}b_{\vec{k},1}+b^{\dagger}_{\vec{k},2}b_{\vec{k},2}).
\end{aligned}
\end{equation}
After the block diagonalization, we can write the spin-angular momentum in each sector and find its average for the magnon mode $\alpha$ and $\beta$ ($\Sigma_{z}=\sigma_{z} \otimes I_{2}$).
\begin{equation}
\begin{aligned}
S_{\uparrow}&=\dfrac{1}{2}\sum_{k}\Psi^{\prime \dagger}_{\uparrow} \Sigma_{z} \Psi^{\prime}_{\uparrow}\\
=&\dfrac{1}{2}\sum_{k}(\alpha^{\dagger}_{\vec{k},1}\alpha_{\vec{k},1}+\beta^{\dagger}_{\vec{k},2}\beta_{\vec{k},2}-\beta_{-\vec{k},1}\beta_{-\vec{k},1}^{\dagger}-\alpha_{-\vec{k},2}\alpha_{-\vec{k},2}^{\dagger}),
\end{aligned}
\end{equation}
and, similarly,
\begin{equation}
\begin{aligned}
S_{\downarrow}&=-\dfrac{1}{2}\sum_{k}\Psi^{\prime \dagger}_{\downarrow} \Sigma_{z} \Psi^{\prime}_{\downarrow}\\
=&-\dfrac{1}{2}\sum_{k}(\alpha^{\dagger}_{\vec{k},2}\alpha_{\vec{k},2}+\beta^{\dagger}_{\vec{k},1}\beta_{\vec{k},1}-\beta_{-\vec{k},2}\beta_{-\vec{k},2}^{\dagger}-\alpha_{-\vec{k},1}\alpha_{-\vec{k},1}^{\dagger}).
\end{aligned}
\end{equation}
In each of these blocks, we calculated the expectation value of the total spin operator which signifies spin-momentum locking of the magnon modes having chirality $\pm 1$, and also, that these expectations are $k$ independent. Interactions like an in-plane easy-axis anisotropy or the Kitaev term destroys this spin-rotation symmetry around $\textit{z}$ axis and invalidates these relations.

\section{Numerical Results}\label{numerical results}

 \begin{figure*}[t]
	\includegraphics[width=1.0\linewidth]{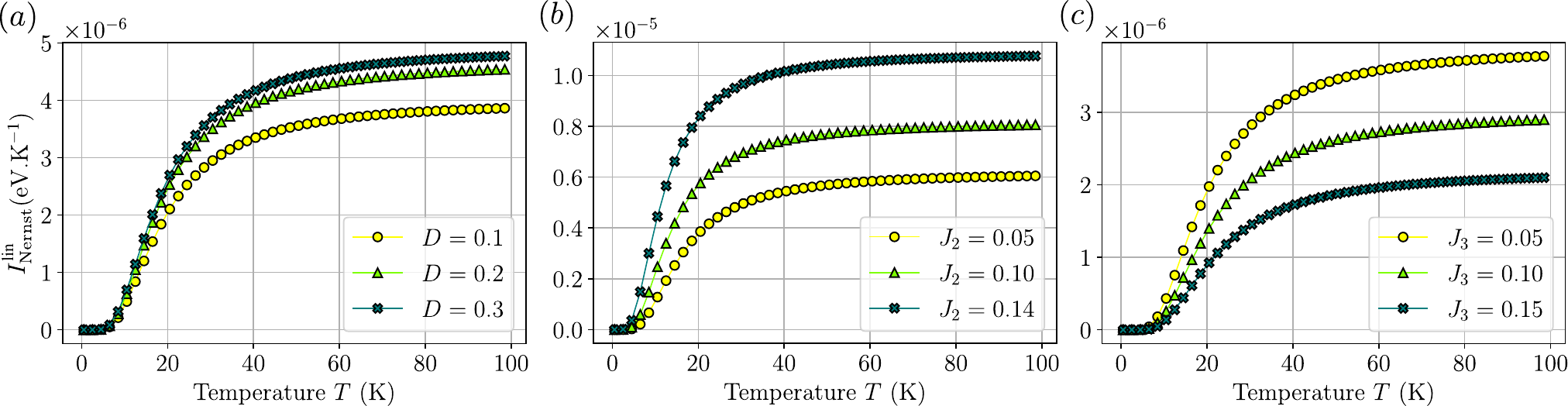}
	\caption{ In Fig (a), (b), and (c) we plot the linear Nernst current as a function of temperature for different values of $D$, $J_{2}$ and $J_{3}$. The coefficients of linear Nernst current is defined in Eq.~(\ref{linearNernstplot}). The Chern number of the bands are $C=\pm 1$ which indicates that the bands are topologically nontrivial. In Fig. (a) $J_{2}$ and $J_{3}$ are kept zero, in Fig. (b) and (c), $D$=0.3 \text{meV}, all the other parameters are the same as Fig.~\ref{fig:lineardnew1}. In a typical experimental setup $\nabla T=10^{-6}$ $\text{K}/ \text{nm}$.}
	\label{fig:lineardnew}
\end{figure*}

\subsubsection{Dispersion and Berry curvature with $D \neq 0$ and $U=0$}
For a single-layer model, with antiferromagnetic Heisenberg interaction in the presence of single ion anisotropy and DMI coupling, the magnon bands are known to be two-fold degenerate ($E_{1}(\vec{k})=E_{2}(\vec{k}), \ E_{1,2}(\vec{k}) \neq  E_{1,2}(\vec{-k})$) with opposite Berry curvature ($\Omega_{1}(\vec{k})=-\Omega_{2}(\vec{k})$). As a consequence, the linear spin-Nernst current becomes non-zero but the thermal Hall current remains zero~\cite{Xioprl}, which can be readily understood from our equations (\ref{nonlinearNersnt}) and (\ref{nonlinearHall2}), respectively. The same model in a bilayer honeycomb lattice with inter-layer antiferromagnetic coupling was also briefly discussed in Ref.~\cite{alexyprl}. We have studied this particular model (Hamiltonian in Eq.~(\ref{eq:Hamiltonian}) with $U=0$) under the additional presence of second and third-nearest-neighbor Heisenberg coupling which was not investigated in earlier literature. In Fig.~\ref{fig:lineardnew1}(a) and \ref{fig:lineardnew1}(b) we plot the magnon spectrum along the high symmetry points with zero and nonzero value of DMI strength. In both cases, the bands are doubly degenerate with vanishing energy at $\Gamma$ point. In contrast to the single layer model, in this case $E_n(\vec{k})=E_n(\vec{-k})$. In absence of DMI, the magnon bands touch at the $K,K'$ points. In Fig.~\ref{fig:lineardnew1}(c) we also show the momentum resolved Berry curvature which peaks near the $M$ points. It is clear that the Berry curvature for this case is an even function of the momentum that results in nonzero value of the Chern number (which is $\pm 1$). In the Table~\ref{tab:Uneq0} we have summarised the symmetries of the dispersion and the Berry curvature for this particular model.

\begin{figure*}[ht]
	\centering
	\includegraphics[width=1.0\linewidth]{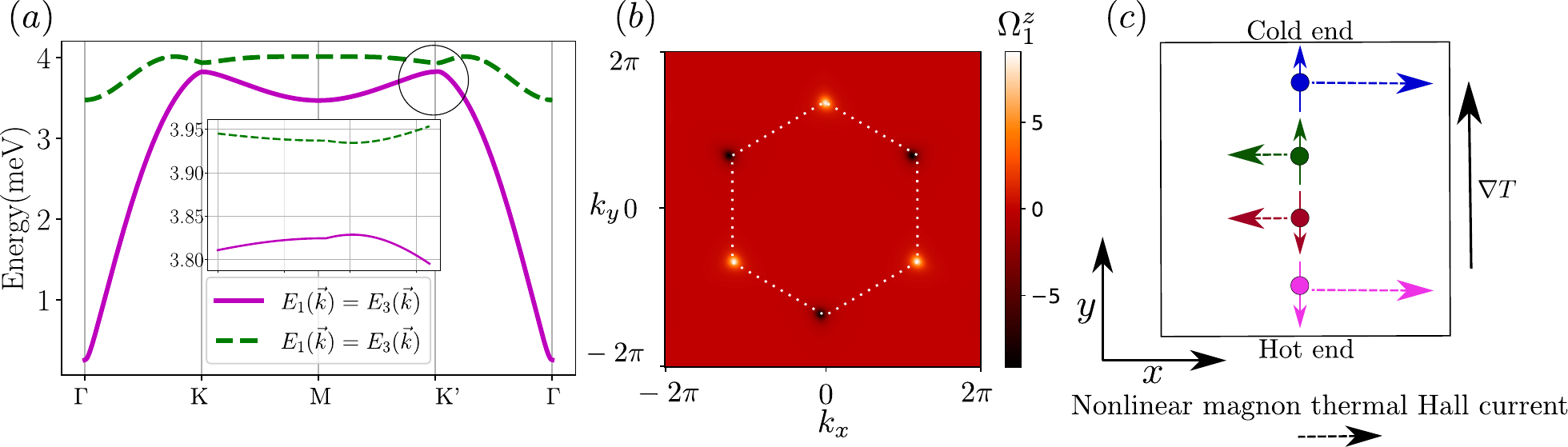}
	\caption{Model with $D=0$: In figures (a) and (b) we plot the band structure along the high symmetry points and the Berry curvature of the lowest magnon band. Parameters of the plots are as follows, $S$=1, $J_{11}=1.0$ \text{meV}, $J_{12}=1.05$ \text{meV} and $J_{13}=0.95$ \text{meV}, $K$=0.0086 \text{meV}, $D$=0, $t$=1 \text{meV}, $U$=0.05 \text{meV}, $J_{2}=0$, $J_{3}=0$. Bands are two-fold degenerate and the gap at $K$, $K'$ points are of the order of twice the ED potential. The Berry curvature for the bands picks up near the $K$ and $K'$ points and is an odd function of momentum, resulting in a zero Chern number. (c) Schematic of nonlinear magnon thermal Hall current. Different colors and the arrows represent different magnon modes and their spin $S^{z}$ quantum numbers, respectively. The length of the arrows from the center represents the magnitude of the corresponding particle current. We have a pair of modes having the same magnitude of Hall current in the same direction but with opposite $S^{z}$, as a result, we have a nonzero nonlinear thermal Hall current with vanishing nonlinear Spin-Nernst current.}
	\label{fig:nonlineararxiv1}
\end{figure*}
\begin{figure*}[ht]
	\centering
	\includegraphics[width=0.8\linewidth]{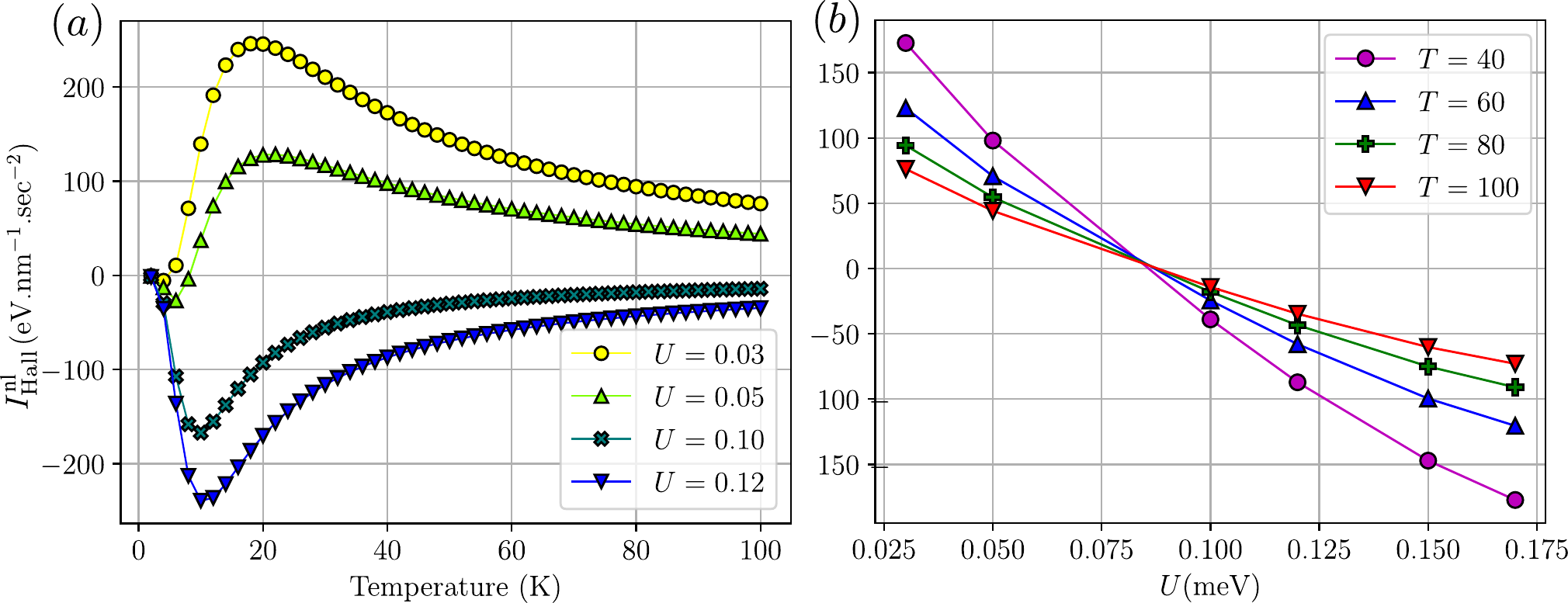}
	\caption{In figures (a) and (b) we plot the nonlinear Hall coefficient defined in Eq.~(\ref{nonlinearHallcurrent}) as a function of temperature and ED potential. We find that, under the presence of strain, there is a sign reversal of the Hall coefficient. For very large anisotropy there is a competition between the Berry curvature density at the $\Gamma$, $M$, and $K$ points. The magnitude of the ED potential should be kept small so that there is no spin-flipping transitions. Other parameters of the plot are same as in Fig.~(\ref{fig:nonlineararxiv1}). The relaxation time ($\tau$) of the magnon modes in antiferromagnets are typically of order $10^{-7}-10^{-9}$ sec.}
	\label{fig:nonlineararxiv2}
\end{figure*}

\begin{table}
\begin{tabular}{ |p{2.cm}||p{2.cm}|p{2.cm}|p{2.cm}|}
	\hline
	Band $n$ & Energy& $\langle S_{z} \rangle$ & $\Omega^{z}_{n}(\vec{k})$\\
	\hline
	1& $E_{1}(\vec{k})$  & +1   & $+\Omega_{1}(\vec{k})$\\
	2& $E_{2}(\vec{k})$  & +1   & $-\Omega_{1}(\vec{k})$\\
	3& $E_{1}(\vec{k})$  & -1   & $+\Omega_{1}(\vec{k})$\\
	4& $E_{2}(\vec{k})$  & -1   & $-\Omega_{1}(\vec{k})$\\	
	\hline
\end{tabular}
	\caption{Model, Eq.~(\ref{eq:Hamiltonian}), with $D=0$ and $U \neq 0$.}
	\label{tab:Uneq0}
\end{table}

\subsubsection{Linear magnon transport}
In this bilayer model, the linear spin-Nernst current is enhanced compared to the single layer. Although the bands are topologically nontrivial, the thermal Hall current remains zero due to a global time-reversal symmetry~\cite{alexyprl}. We neglect the nonlinear part of the Eq.~(\ref{nonlinearNersnt}) and write down $J_x^{\rm lin, \rm Nernst}=I^{\rm lin}_{\rm Nernst} \nabla T$ (see details in Appendix A, Eq.~(\ref{linearNernstplot})). In Fig.~\ref{fig:lineardnew} we plot $I^{\rm lin}_{\rm Nernst}$ as a function of temperature for different values of DMI strength, as well as, $J_{2}$ and $J_{3}$. For the increasing value of $D$ the linear spin-Nernst current increases. As a function of temperature, it starts from zero, then increases, and finally saturates.
As we increase $J_{2}$ the nearest neighbor spins get frustrated, which helps non-collinear configurations, in contrast, $J_{3}$ stabilizes the N\'{e}el ordering. This is the reason why for constant $D$, the linear Nernst coefficient increases with increasing $J_{2}$ but decreases with $J_{3}$. From the perspective of  magnon dispersion, with increasing $J_{2}$ the inter-band gaps between the magnon bands at $M$ points increase, in contrast, when $J_{3}$ is increased the gaps decrease, leading to a vanishing measure of the Berry curvature at the $M$ points, and, as a result, the linear Nernst current decreases with increasing $J_{3}$. In this analysis, we have kept the values of $J_{2}$ to be small enough so that the system is still in an ordered state and the spin-wave theory is a valid approximation~\cite{mukherjee2021schwinger}. Our analysis reveals that the change in magnitude of the linear spin-Nernst coefficients by varying second and third nearest neighbor Heisenberg coupling is much larger in comparison to the change due to $D$. In passing we comment that the nonzero spin-Nernst current observed in the material $\ce{MnPS_{3}}$ was originally explained using these models, but recent neutron experiments done on the same material~\cite{PhysRevB.103.024424} suggests that the observed value of $D$ is too small to explain the magnitude of linear spin-Nernst current. Thus, the observed Nernst effect may be related to other possible mechanisms, such as the magnon-magnon and magnon-phonon interaction~\cite{PhysRevLett.123.167202} and nonlinear effects.

\subsubsection{Dispersion and Berry curvature with $U \neq 0$ and $D=0$}
In a very recent work~\cite{PhysRevResearch.4.013186} authors have shown that in a single layer honeycomb lattice, even without DMI, in the presence of anisotropic Heisenberg exchange interaction, one can get nonvanishing magnon spin-Nernst current.
In this case, the dispersion and Berry curvature holds the following identities $E_{1}(\vec{k})=E_{2}(\vec{k}), \ E_{1,2}(\vec{k})=E_{1,2}(\vec{-k})$ and $\Omega_{1}(\vec{k})=-\Omega_{2}(\vec{k})$, because of this symmetry, the linear spin-Nernst and thermal Hall current remains zero. 
In following, we investigate the nonlinear response in a stacked bilayer honeycomb lattice by introducing a layer-dependent electrostatic potential that can be externally controlled by changing the amount of doping~\cite{he2010robust,ahn2006electrostatic,jiang2018controlling}. The Hamiltonian is given by Eq.~(\ref{eq:Hamiltonian}) with $D=0$ and in addition, we have a strain-induced anisotropic nearest neighbor coupling: $J_{11}\neq J_{12} \neq J_{13}$.

Now without the application of strain (i.e, when $J_{11}= J_{12} = J_{13}$), near the $K, K'$ points the derivatives of magnon dispersions are vanishingly small, making their product $\Omega_{n}(\Vec{k})\partial E_n(\vec{k}) / \partial k_{y}$ almost zero near each of those points. However, under the application of strain, the maximum value of Berry curvature and the derivative of dispersion  shifts in the $k_{x}-k_{y}$ plane in a nonequivalent way that makes their product nonzero. This is a necessary condition to get a large nonvanishing nonlinear response in our particular model. Such anisotropic nearest neighbor coupling can be generated by the application of external pressure-induced strain~\cite{PhysRevResearch.4.013186}. As a passing comment, we want to mention that, very high value of ED potential leads to a transition from an antiferromagnetic to a ferromagnetic interlayer coupling even in zero magnetic field~\cite{jiang2018controlling}, so we assume ED potential to be small enough so that the interlayer interaction remains antiferromagnetic in nature. In the Table~\ref{tab:Dneq0} we have summarised the symmetries of the dispersion, Berry curvature for this particular model. 

In Fig.~\ref{fig:nonlineararxiv1} we plot the magnon dispersions and the Berry curvature for this model with $J_{11}\neq J_{12} \neq J_{13}$. The spectrum is doubly degenerate where the gaps between the bands at $K,K'$ points are proportional to $U$. In this model, the maximum contributions to the Berry curvature come from momenta near the $K$, $K'$ points. The Berry-curvature in this case being an odd function of the  Bloch momentum, the Chern number of the band is zero. 

\vspace{.5cm}
\begin{table}
\begin{tabular}{ |p{2.cm}||p{2.cm}|p{2.cm}|p{2.cm}|}
	\hline
	Band $n$ & Energy& $\langle S_{z} \rangle$ &$\Omega^{z}_{n}(\vec{k})$\\
	\hline
	1& $E_{1}(\vec{k})$  & +1   & $+\Omega_{1}(\vec{k})$\\
	2& $E_{2}(\vec{k})$  & +1   & $-\Omega_{1}(\vec{k})$\\
	3& $E_{1}(\vec{k})$  & -1   & $-\Omega_{1}(\vec{k})$\\
	4& $E_{2}(\vec{k})$  & -1   & $+\Omega_{1}(\vec{k})$\\	
	\hline
\end{tabular}
\caption{ Model, Eq.~(\ref{eq:Hamiltonian}), with $U=0$ and $D \neq 0$. In this case the two bands have Chern numbers $C=\pm 1$.}
\label{tab:Dneq0}
\end{table}

\begin{figure*}[ht]
	\centering
	\includegraphics[width=1.0\linewidth]{"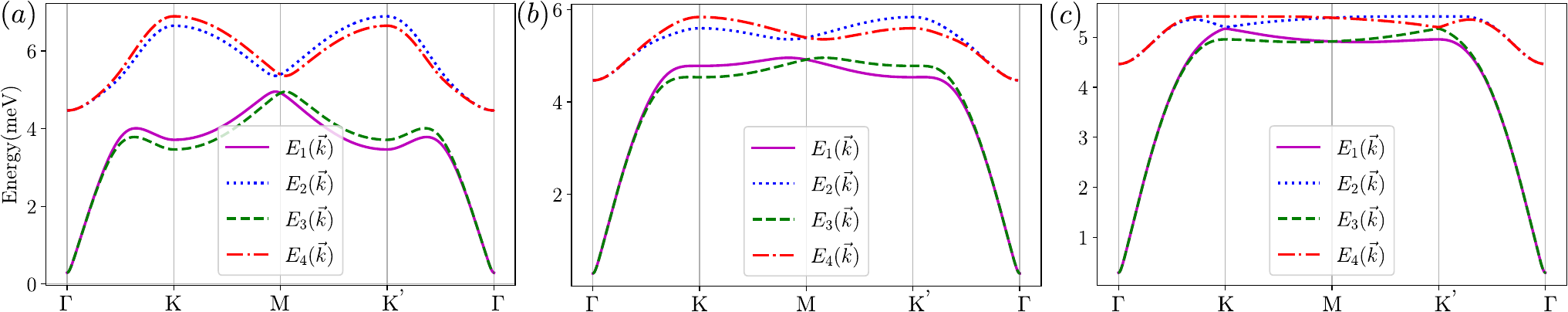"}
	\caption{In this figure we show the evolution of  magnon band structures in the $D+U$ model with a varying magnitude of $D$ and $U$. Parameters are as follows: (a) $U/D=0.4$, (b) $U/D=1.2$ (c) $U/D=6.0$, all the other parameters are same as Fig.~\ref{fig:lineardnew}.  It is interesting to note that, in $D+U$ model the valley degeneracy between $K,K'$ points is broken, i.e, the magnon bands have different energies at those two momentum points.}
	\label{fig:dandu-new}
\end{figure*}
\begin{table*}[ht]
	\begin{tabular}{ |p{5.5cm}||p{2.8cm}|p{2.8cm}|p{2.8cm}|p{2.8cm}|}
		\hline
		Model & Linear Nernst& Linear Hall  & Nonliner Nernst &Nonlinear Hall\\
		\hline
		$U$ $\neq$ 0, $D$=0 & \textcolor{red}{\xmark} & \textcolor{red}{\xmark}  & \textcolor{red}{\xmark}&  \textcolor{red}{\xmark} \\
		$U$ $\neq$ 0, $D$=0 +$\text{Strain}$& \textcolor{red}{\xmark}  & \textcolor{red}{\xmark}   & \textcolor{red}{\xmark}&  \textcolor{blue}{\checkmark}\\
		$D$ $\neq$ 0, $U$=0& \textcolor{blue}{\checkmark}  & \textcolor{red}{\xmark}  & \textcolor{red}{\xmark} &  \textcolor{red}{\xmark}\\
		\hline
	\end{tabular}
	\caption{Varients of $D+U$ model and their various transport signatures.}
	\label{tab:DU}
\end{table*}

\subsubsection{Nonlinear magnon transport}
In our case, the inversion symmetry is broken in each layer but it remains intact if we consider both the  layers together. 
From Eq.~(\ref{nonlinearNersnt}), the total nonlinear magnon spin-Nernst current can be written as, $$J_{x}^{\rm nl,\rm Nernst}=\dfrac{\tau (\nabla T)^2}{\hbar VT^2}\sum_{n,\vec{k}}  \langle S_{n}^{z} \rangle \Omega_{n}(\vec{k}) g_{1}(E_{n}(\vec{k})),$$ where, $g_{1}(E_{n}(\vec{k}))=E_{n}(\vec{k})^2 \partial \rho^{(0)}_{n}/ \partial k_{y} $ and the Berry curvature $\Omega_{n}(\vec{k})$ are both odd functions of momentum. As a result, the nonlinear Nernst response for the individual magnon bands will be nonzero but when we sum over the bands they cancel each other out.
It means that we have counter-propagating nonlinear spin currents. 

Interestingly, the nonlinear thermal Hall response is nonzero in this case. From the Eq.~(\ref{nonlinearHall2}), the total nonlinear magnon thermal Hall current can be expressed as,
$$J_{x}^{\rm nl, \rm Energy}=\dfrac{\tau (\nabla T)^2}{\hbar^2 VT^2} \sum_{n,\vec{k}} \Omega_{n}(\vec{k}) g_{2}(E_{n}(\vec{k})),$$ here, $\Omega_{n}(\vec{k})$ and $g_{2}(E_{n}(\vec{k}))=E_{n}(\vec{k})^3 \partial \rho^{(0)}_{n}/ \partial k_{y}$ are both odd under $k$.  The nonlinear magnon thermal Hall response for each bands as well as thier sum is nonzero. In Fig.~\ref{fig:nonlineararxiv2}(a) we  plot the nonlinear magnon thermal Hall current (details in the Appendix A, Eq.~(\ref{nonlinearHallcurrent})) as a function of temperature for different values of $U$. The thermal Hall  coefficient starts from zero and peaks up at a point where the temperature becomes of the order of the energy gap. From Fig.~\ref{fig:nonlineararxiv2}(b), it is also clear that with increasing $U$ there is a sign change in the nonlinear Hall current. These results predict that the nonlinear thermal Hall current can indeed be tuned by external doping and strain-induced anisotropy. These are the main results of our current work.

In our nonlinear Hall response, most of the contribution comes near the $K,K'$ points, near which the group velocities of magnons are of the order $\dfrac{1}{\hbar}\dfrac{\partial E}{\partial k}=7.5\times$ $10^{11}$ nm sec$^{-1}$. Typically, the magnon mean free path for an antiferromagnetic sample at 20 $K$ ranges from 1-100 $\mu m$~\cite{PhysRevB.90.064421}, this corresponds to a magnon lifetime ($\tau$) is $10^{-7}-10^{-9}$ seconds. The applied temperature gradient ($\nabla T$) for a magnon transport measurement reported by the experiment in Ref.~[\onlinecite{PhysRevB.96.134425}] is of order of $10^{-6}$ K/nm. The coefficient of nonlinear Hall current we obtain for $U=0.03$ meV at 20K is around 250 eV nm$^{-1}$ sec$^{-2}$. This is equivalent to a nonlinear thermal Hall current of 250$\times$10$^{-7}$ eV~nm$^{-1}$sec$^{-1}$ $\approx 10^{-14}$ W/m, which is in the measurable range. In comparison, the magnitude of the linear magnon thermal Hall conductivity reported in Ref.~\cite{PhysRevLett.123.167202,doi:10.1126/science.1188260} at 20 K is around $10^{-13}$ W/K. Assuming the same value of the temperature gradient in our case, the value of the linear magnon Thermal Hall current is $10^{-10}$ W/m. We have checked that the order of magnetude estimation is robust against changes of the material parameters.
\subsubsection*{Model with both $U \neq 0$ and $D \neq 0$}
We have also analyzed the magnon band structure when both DMI and ED are nonzero~(Fig.~\ref{fig:dandu-new}). For this particular case, the two-fold degeneracy of the magnon modes is lifted and the dispersion becomes asymmetric about the $\Gamma$ point (also termed as  nonreciprocal magnons) with the formation of Dirac-like nodes near the $M$ point. The degree of non-reciprocity of the magnon band structures can be possibly manipulated by changing the direction and magnitude of the external ED potential. As the linear response is already non-zero for this model, we do not show any transport studies of this model in this paper. The outcome of various magnon transport coefficients for variants of $U+D$ model are summarized in the Table~\ref{tab:DU}.

\subsubsection*{Momentum and temperature resolved relaxation time}
The simplest mechanism through which out-of-equilibrium magnets can relax is known as Gilbert damping~\cite{gilbert1,gilbert2,gilbert3,gilbert4}.  From the Landau-Lifshitz-Gilbert equations, the scattering rate can be written as,
\begin{equation}
    \Gamma=\dfrac{\partial \rho}{\partial t}=-\dfrac{1}{\tau_{G}}(\rho_{\Vec{k}}-\rho_{eq})=-\dfrac{2\alpha E(\Vec{k}) }{\hbar}(\rho_{\Vec{k}}-\rho_{eq}),
    \end{equation}
where $\rho$ is the bosonic distribution function, $\alpha$ is the Gilbert damping parameter and $E(\vec{k})$ is the magnon dispersion. In this mechanism the relaxation time is inversely proportional to the dispersion, as a result, we can expect that at small temperatures, it will modify the magnitude of the different magnon transport coefficients. For example, within the Gilbert relaxation, our nonlinear magnon Hall current will be proportional to $E(\Vec{k})^2$ instead of $E(\Vec{k})^3$. For higher temperatures, magnon-magnon interactions become important and it can significantly modify the band structures and the wavefunctions. Previous works~\cite{rel1,rel2,rel3} have confirmed that there is a $T^2$ dependence on relaxation rates. In our case, this will make the nonlinear Hall current proportional to $T^{0}$ in contrast to $T^{-2}$ dependence under constant relaxation time. This certainly enhances the magnitude of the Hall response at higher temperatures.

\section{Summery}\label{discussion}
In conclusion, we have investigated the linear and nonlinear magnon transport under the presence of various possible spin-spin interactions in a bilayer van der Waals honeycomb magnet within the semiclassical Boltzmann transport theory. We have shown that, even in the absence of Dzyaloshinskii-Moriya interactions (DMI), the presence of anisotropy and electrostatic doping potential (ED) can lead to a nonzero \textit{nonlinear} thermal Hall effect. Interestingly, we have observed a sign reversal of this nonlinear magnon Hall current as a function of the ED potential which can have the potential for application in spin-based technologies. We have further shown that, in the presence of DMI coupling, the second and third nearest Heisenberg interactions play an important role in determining the magnitude of the \textit{linear} magnon spin-Nernst current. We have also commented on the momentum and Temperature dependence of the magnon scattering time which can significantly affect the magnitude of the transport coefficients and their experimental relevance.

\section{ACKNOWLEDGMENTS}
R.M and S.V thanks for useful communication with Ran Cheng (University of California, Riverside), Vladimir A. Zyuzin (Landau Institute, Moscow), Hiroki Kondo (University of Tokyo), Yutaka Akagi (University of Tokyo). R.M. acknowledges the CSIR (Govt. of India) for financial support. We also acknowledge the use of the HPC facility at IIT Kanpur. A.K  acknowledges support from the SERB (Govt. of India) via saction no. ECR/2018/001443 and CRG/2020/001803, DAE (Govt. of India ) via sanction no. 58/20/15/2019-BRNS, as well as MHRD (Govt. of India) via sanction no. SPARC/2018-2019/P538/SL.

\bibliography{nonlinearv2.bib}

	\renewcommand{\theequation}{S\arabic{equation}}
\setcounter{equation}{0}
\renewcommand{\thefigure}{S\arabic{figure}}
\setcounter{figure}{0}
\appendix{}
\begin{widetext}
	
\section*{Appendix A : Semiclassical Boltzmann-Transport calculation}
The semiclassical equations of motion of the magnon Bloch bands are given by,
\begin{equation}\label{semiclass1}
\dot{\vec{r}}=\dfrac{1}{\hbar}\dfrac{\partial E_n(\vec{k})}{\partial \vec{k}}- \dot{\vec{k}} \times \vec{\Omega}_{n}(\vec{k}) ,
\end{equation}
\begin{equation}\label{semiclass2}
\hbar \dot{\vec{k}}= - \vec{\nabla} V_{\text{con}}(\vec{r}),
\end{equation}
here $n$ is the band index, $E_n(\vec{k})$ is the $n^{th}$ magnon band energy, $\Omega_{n}^{z}(\vec{k})$ is the Berry curvature in momentum space. For the validity of Eq.~(\ref{semiclass1}) and Eq.~(\ref{semiclass2}), the spatial variation of the confining potential $V_{\text{con}}(\vec{r})$ should be much slower compared  with the size of the magnon wave packet. Here, we focus on the edge current in the $x$-direction, with a small temperature gradient in the $y$-direction as an example.

We are specifically interested in the situation when the contribution due to the first term of Eq.~(\ref{eq:currentd}) (which is linear in $\Delta T$) vanishes due to symmetry considerations. We show below that the second term is proportional to $(\Delta T)^2$, which gives the first order nonlinear correction.
\begin{equation}\label{currenttot}
j_{n,x}^{\rm nl}(y)= \dfrac{1}{V}\sum_{\vec{k}} \rho_{n}^{(1)}(\vec{k};T(y))\dfrac{1}{\hbar} \dfrac{dV_{\text{con}}(y)}{dy} \Omega^{z}_{n}(\vec{k}).
\end{equation}
Under the relaxation time approximation, as written in the main text,
\begin{equation*}
\dot{\vec{r}} \cdot \dfrac{\partial \rho}{\partial r}+\dot{\vec{k}} \cdot \dfrac{\partial \rho}{\partial k}= -\dfrac{(\rho-\rho^{(0)})}{\tau},
\end{equation*}
where $\rho^{(0)}$ is the equilibrium distribution function. Let us suppress the suffix $n$ in $\rho_{n}$ for notational simplicity for the moment. We first calculate the first order correction $ie$. $\rho^{(1)}$
\begin{equation*}
\dot{\vec{r}} \cdot \dfrac{\partial\rho^{(0)}}{\partial r}+\dot{\vec{k}} \cdot \dfrac{\partial\rho^{(0)}}{\partial k}= -\dfrac{\rho^{(1)}}{\tau} \ \ \ \ \Big(\dfrac{\partial \rho^{(0)}}{\partial x}=0, \ \ \ \dfrac{\partial \rho^{(0)}}{\partial y} \neq 0 \Big)
\end{equation*} 
using Eq.~(\ref{boltz11}) and Eq.~(\ref{boltz22}) we can write,
\begin{equation}\label{eq:49}
\begin{aligned}
&v_{y}\dfrac{\partial \rho^{(0)}}{\partial y}-\dfrac{1}{\hbar} \dfrac{dV_{\text{con}}}{dy}\dfrac{\partial \rho^{(0)}}{\partial k_ {y}}=\dfrac{-\rho^{(1)}}{\tau}\\
\Rightarrow &\dfrac{1}{\hbar}\dfrac{\partial E_n(\vec{k})}{\partial k_{y}}\dfrac{\partial \rho^{(0)}}{\partial y}-\dfrac{1}{\hbar} \dfrac{dV_{\text{con}}}{dy}\dfrac{\partial \rho^{(0)}}{\partial k_ {y}}=\dfrac{-\rho^{(1)}}{\tau}.
\end{aligned}
\end{equation} 
And we write,
\begin{equation}
\dfrac{\partial \rho^{(0)}}{\partial y}=\dfrac{\partial \rho^{(0)}}{\partial T} \dfrac{dT}{dy} +\dfrac{\partial \rho^{(0)}}{\partial V_{\text{con}}} \dfrac{dV_{\text{con}}}{dy}. 
\end{equation}
From Eq~(\ref{eq:49}), we can write,
\begin{equation}
\dfrac{\partial E_n(\vec{k})}{\partial k_{y}}\dfrac{\partial \rho^{(0)}}{\partial T}\dfrac{dT}{d y}+
\dfrac{\partial E_n(\vec{k})}{\partial k_{y}}\dfrac{\partial \rho^{(0)}}{\partial V_{con}}\dfrac{dV_{\text{con}}}{d y}-\dfrac{dV_{\text{con}}}{dy}\dfrac{\partial \rho^{(0)}}{\partial k_{y}}=-\dfrac{\hbar}{\tau} \rho^{(1)}.
\end{equation}
The equilibrium bosonic distribution is given by,
\begin{equation*}
\rho^{(0)}=\dfrac{1}{e^{\beta(E_n(\vec{k})-\mu)}- 1}
\end{equation*}
\begin{equation*}
\Rightarrow \dfrac{\partial \rho^{(0)}}{\partial T}=\dfrac{-1}{\big(e^{\beta(E_n(\vec{k})-\mu)}- 1\big)^{2}}\Big(-\dfrac{E_n(\vec{k})-\mu}{k_{B}T^{2}}\Big), \ \ \dfrac{\partial \rho^{(0)}}{\partial E_n(\vec{k})}=\dfrac{-1}{\big(e^{\beta(E_n(\vec{k})-\mu)}- 1\big)^{2}}\Big( \dfrac{1}{k_{B}T}   \Big).
\end{equation*}
So we have,
\begin{equation*}
\dfrac{\partial \rho^{(0)}}{\partial T}=\Big(-\dfrac{E_n(\vec{k})-\mu}{T} \Big)\dfrac{\partial \rho^{(0)}}{\partial E_n(\vec{k})},
\end{equation*}
this expression enables us to write $\rho^{(1)}$ in compact notation,
\begin{equation}\label{rho1}
\rho^{(1)}=\dfrac{-\tau}{\hbar}\Big(-\dfrac{E_n(\vec{k})-\mu}{T}\Big)\dfrac{\partial E_n(\vec{k})}{\partial k_{y}}
\dfrac{\partial \rho^{(0)}}{\partial E_n(\vec{k})}\dfrac{dT}{d y}-\dfrac{\tau}{\hbar}\dfrac{\partial E_n(\vec{k})}{\partial k_{y}}\dfrac{\partial \rho^{(0)}}{\partial V_{\text{con}}}\dfrac{dV_{\text{con}}}{dy}+\dfrac{\tau}{\hbar}\dfrac{dV_{\text{con}}}{dy}\dfrac{\partial\rho^{(0)}}{\partial k_{y}}.
\end{equation}
Now we neglect the second and third contributions arising in Eq.~(\ref{rho1}) as they will give rise to contributions which are of higher order in $(\Delta T)^2$. Under the assumption:
\begin{equation*}
\Big (\dfrac{dT}{dy} \Big ) \gg \Big( \dfrac{dV_{\text{con}}}{dy}\Big),
\end{equation*}
we have,
\begin{equation}
\rho^{(1)}(\vec{k},T(y))=\dfrac{\tau}{\hbar}\dfrac{E_n(\vec{k})-\mu}{T}\dfrac{\partial \rho^{(0)}}{\partial k_{y}} \left(\dfrac{dT}{dy}  \right).
\end{equation}
The expression of nonlinear current density for each band can be written as (puting back the suffix $n$),

\begin{equation*}
j_{n,x}^{\rm nl}(y)= \dfrac{1}{V} \sum_{\vec{k}}  \dfrac{1}{\hbar} \dfrac{dV_{\text{con}}(y)}{dy} \Omega^{z}_{n}(\vec{k})\dfrac{\tau}{\hbar}\dfrac{E_n(\vec{k})-\mu}{T}\dfrac{\partial\rho_{n}^{(0)}}{\partial k_{y}} \left(\dfrac{dT}{dy}  \right).
\end{equation*}
The total averaged nonlinear current in $x$ direction is given by the integral of current density
\begin{equation*}
J^{\rm nl}_{n,x}= \dfrac{1}{V} \sum_{\vec{k}}  \dfrac{1}{\hbar} \Omega^{z}_{n}(\vec{k})\dfrac{\tau}{\hbar}\dfrac{E_n(\vec{k})-\mu}{T} \left(\dfrac{dT}{dy}  \right)  \int_{0}^{\infty}\dfrac{1}{w}\Big(\dfrac{\partial \rho_{n}^{(0)}(E_n(\vec{k})+V_{\text{con}}(r);T(+w/2))}{\partial k_{y}}-\dfrac{\partial \rho_{n}^{(0)}(E_n(\vec{k})+V_{\text{con}}(r);T(-w/2))}{\partial k_{y}} \Big)\  dV_{\text{con}} 
\end{equation*}
This is zero if $T(w/2)=T(-w/2)$. Now using Taylor series approximation we can write,
\begin{equation*}
\rho_{n}^{(0)}(T(-y))=\rho_{n}^{(0)}(T(y))-2y\dfrac{dT}{dy}\dfrac{\partial \rho_{n}^{(0)}}{\partial T}.
\end{equation*}
Thus,
\begin{equation*}
J^{\rm nl}_{n,x}= \dfrac{1}{V} \sum_{\vec{k}}  \dfrac{1}{\hbar} \Omega^{z}_{n}(\vec{k})\dfrac{\tau}{\hbar}\dfrac{E_n(\vec{k})-\mu}{T} \left(\dfrac{dT}{dy}  \right)^{2}  \int_{0}^{\infty}\Big(\dfrac{\partial^{2} \rho_{n}^{(0)}(E_n(\vec{k})+V_{\text{con}}(r))}{\partial k_{y} \partial T} \Big)\  dV_{\text{con}}.
\end{equation*}
We have,
\begin{equation*}
\dfrac{\partial \rho_{n}^{(0)}}{\partial k_{y}}=\dfrac{\partial \rho_{n}^{(0)}(E_n(\vec{k})+V_{\text{con}}(r))}{\partial E_n(\vec{k})}\dfrac{\partial E_n(\vec{k}) }{\partial k_{y}},
\end{equation*}
and,
\begin{equation}
\begin{aligned}
J^{\rm nl}_{n,x}&= \dfrac{1}{V} \sum_{\vec{k}}  \dfrac{1}{\hbar} \Omega^{z}_{n}(\vec{k})\dfrac{\tau}{\hbar}\dfrac{E_n(\vec{k})-\mu}{T} (\nabla T)^{2} \dfrac{\partial}{\partial T}  \int_{0}^{\infty}\Big(\dfrac{\partial \rho_n^{(0)}(E_n(\vec{k})+V_{\text{con}}(r))}{\partial k_{y}} \Big)\  dV_{\text{con}}\\
&=\dfrac{\tau(\nabla T)^{2}}{\hbar T} \sum_{\vec{k}} \dfrac{1}{V}  \dfrac{1}{\hbar} \Omega^{z}_{n}(\vec{k})\dfrac{(E_n(\vec{k})-\mu)^{2}}{T}  
\dfrac{1}{k_{B}T}\rho^{(0)}(E_n(\vec{k}))\Big(1+\rho^{(0)}(E_n(\vec{k}))\Big)\dfrac{\partial E_n(\vec{k}) }{\partial k_{y}} .
\end{aligned}
\end{equation}
Total nonlinear spin-Nernst current is given by,
\begin{equation}\label{nonlinear_Nernst_tot}
J_{x}^{\rm nl,\rm Nernst}=\hbar \sum_{n} \langle S^{z}_{n} \rangle J_{n,x}^{\rm nl}.
\end{equation}
The total averaged magnon current for each band including both linear~\cite{Xioprl,PhysRevB.84.184406} and nonlinear contribution is given by,
\begin{equation}
J_{n,x}=\dfrac{k_{B}}{V} \sum_{\vec{k}}  \dfrac{1}{\hbar} \Omega^{z}_{n}(\vec{k})c_{1}(\rho^{(0)}_{n})(\nabla T)+ \dfrac{1}{V} \sum_{\vec{k}}  \dfrac{1}{\hbar} \Omega^{z}_{n}(\vec{k})\dfrac{\tau}{\hbar}\dfrac{(E_n(\vec{k})-\mu)^{2}}{T^{2}}\dfrac{\partial \rho^{(0)}_{n}}{\partial k_{y}}(\nabla T)^{2},
\end{equation}
the first term is the linear contribution,
\begin{equation}
J^{\rm lin}_{n,x}=\dfrac{k_{B}}{V} \sum_{\vec{k}}  \dfrac{1}{\hbar} \Omega^{z}_{n}(\vec{k})c_{1}(\rho^{(0)}_{n})(\nabla T).
\end{equation}
The linear spin-Nernst current is given by,
\begin{equation}\label{linearNernstplot}
\begin{aligned}
J_{x}^{\rm lin,\rm Nernst}&=\hbar \sum_{n} \langle S^{z}_{n} \rangle J_{n,x}^{\rm lin}\\
&=I^{\rm lin}_{\rm Nernst} \ \nabla T
\end{aligned}
\end{equation}
nonlinear energy current is simply given by,
\begin{equation*}
J^{\rm nl, \rm Energy}_{n,x}= \dfrac{1}{V} \sum_{\vec{k}}   \Omega^{z}_{n}(\vec{k})\dfrac{\tau}{\hbar^{2}}\dfrac{(E_n(\vec{k})-\mu)^{3}}{k_{B}T^{3}}  
\rho^{(0)}(E_n(\vec{k}))\Big[1+\rho^{(0)}(E_n(\vec{k}))\Big]\dfrac{\partial E_n(\vec{k}) }{\partial k_{y}} (\nabla T)^{2},
\end{equation*}
the total nonlinear Hall current is given by,
\begin{equation}\label{nonlinearHallcurrent}
J_{x}^{\rm nl, \rm Energy}=\sum_{n} J^{\rm nl, \rm Energy}_{n,x}=\tau \times I^{\rm nl}_{\rm Hall}.
\end{equation}

\section*{Appendix B: Further details of model}
In this section, we provide the details of different kinds of spin-spin interaction under the linear spin-wave approximation. We calculate the Heisenberg coupling up to the third order, Dzyaloshinskii-Moriya (DM) coupling in second order (first order term is zero in honeycomb lattice from symmetry consideration), and In-plane easy-axis anisotropy term. We write the terms in a symmetrized fashion.

Real space lattice unit vectors of the honeycomb lattice are given by (see Fig.~\ref{fig:highsymmetry}),
\begin{equation}
\vec{a_{1}}=\dfrac{a}{2} \big(3,\sqrt{3}\big), \ \ \vec{a_{2}}=\dfrac{a}{2} \big(3,-\sqrt{3}\big).
\end{equation}
In following, we set the nearest-neighbor spacing $a=1/\sqrt{3}$. The re-scaled nearest-neighbor lattice vectors are then
\begin{equation}
\vec{\delta}_{1}=\dfrac{1}{2}\big(\dfrac{1}{\sqrt{3}},1\big), \ \ \vec{\delta}_{2}=\dfrac{1}{2}\big(\dfrac{1}{\sqrt{3}},-1\big), \ \ 
\vec{\delta}_{3}=\dfrac{1}{\sqrt{3}}\big(-1,0\big).
\end{equation}

The nearest neighbor terms ($H_{H}^{(1)}$) without any anisotrpy ($J_{11}=J_{12}=J_{13}$) is written in the momentum space as,	\begin{flalign}
	H_{H}^{(1)}=\dfrac{J_{1}}{2}\sum_{\vec{k},\vec{\delta}_{i},{i=1,2,3}} (e^{-i\vec{k} \cdot \vec{\delta}_{i}} a_{\vec{k}}b_{-\vec{k}}+ e^{i\vec{k} \cdot \vec{\delta}_{i}} a_{-\vec{k}}b_{\vec{k}}+h.c)+\dfrac{J_{1}}{2}z_{1}\sum_{k}(a^{\dagger}_{\vec{k}}a_{\vec{k}}+a^{\dagger}_{-\vec{k}}a_{-\vec{k}}+b^{\dagger}_{\vec{k}}b_{\vec{k}}+b^{\dagger}_{-\vec{k}}b_{-\vec{k}}),
	\end{flalign}
 where $\vec{\delta}_{i}$ with $i=1,2,3$ are the three nearest neighbor lattice vectors connecting $A$ and $B$ sublattices, and the coordination number, $z_{1}=3$ for honeycomb lattice. In the case of anisotropic interaction $J_{1}z_{1}$ is replaced by respective coupling strength.  Next nearest neighbor Heisenberg interaction ($H_{H}^{(2)}$) is given by,	
	\begin{flalign}
	H_{H}^{(2)}=&\dfrac{J_{2}S}{2}\sum_{\vec{k},\vec{\eta}_{i},{i=1,2,3}} (e^{-i\vec{k} \cdot \vec{\eta}_{i}}a^{\dagger}_{\vec{k}}a_{\vec{k}}+e^{i\vec{k} \cdot \vec{\eta}_{i} }a^{\dagger}_{-\vec{k}}a_{-\vec{k}}+e^{-i\vec{k} \cdot \vec{\eta}_{i} }b^{\dagger}_{\vec{k}}b_{\vec{k}}+e^{i\vec{k} \cdot \vec{\eta}_{i} }b^{\dagger}_{-\vec{k}}b_{-\vec{k}}+h.c)-J_{2}Sz_{2}\sum_{k}(a^{\dagger}_{\vec{k}}a_{\vec{k}}\\ 
	&+a^{\dagger}_{-\vec{k}}a_{-\vec{k}}+b^{\dagger}_{\vec{k}}b_{\vec{k}}+b^{\dagger}_{-\vec{k}}b_{-\vec{k}}), \nonumber
	\end{flalign}
where $\vec{\eta}_{i}$ with $i=1,2,3$ are the three next-nearest neighbor lattice vectors connecting the $AA$ and $BB$ sublattices, $z_{2}$=6 for honeycomb lattice. Third nearest neighbor Heisenberg interaction ($H_{H}^{3}$) is given by,	
	\begin{equation}
	H_{H}^{(3)}=\dfrac{J_{3}S}{2}\sum_{\vec{k},\vec{\zeta}_{i},{i=1,2,3}} (e^{-i\vec{k} \cdot \vec{\zeta}_{i}} a_{\vec{k}}b_{-\vec{k}}+ e^{i\vec{k} \cdot \vec{\zeta}_{i}} a_{-\vec{k}}b_{\vec{k}}+h.c)+\dfrac{J_{3}S}{2}z_{3}\sum_{k}(a^{\dagger}_{\vec{k}}a_{\vec{k}}+a^{\dagger}_{-\vec{k}}a_{-\vec{k}}+b^{\dagger}_{\vec{k}}b_{\vec{k}}+b^{\dagger}_{-\vec{k}}b_{-\vec{k}}),
	\end{equation}
 where $\vec{\zeta}_{i}$ with $i=1,2,3$ are the three third-nearest neighbor lattice vectors connecting $AB$ sublattices, $z_{3}$=3 for honeycomb lattice. The easy axis anisotropy term which stabilized the ordering along the c axis is given by ($H_{E}$),	
	\begin{equation}
	\begin{aligned}
	H_{E}&=\sum_{i}K (S_{i}^{z})^{2}=\dfrac{-(2S-1)KS}{2S}\sum_{k}(a^{\dagger}_{\vec{k}}a_{\vec{k}}+a^{\dagger}_{-\vec{k}}a_{-\vec{k}}+b^{\dagger}_{\vec{k}}b_{\vec{k}}+b^{\dagger}_{-\vec{k}}b_{-\vec{k}})
	\end{aligned}
	\end{equation}
and the DMI coupling term ($H_{\rm DM}$)	between next nearest neibour is given by,
	\begin{equation}
	\begin{aligned}
	H_{\rm DM}=&\sum_{i,j} \nu_{ij} D \hat{z} \cdot (\vec{S}_{i} \times \vec{S}_{j})=\dfrac{S}{2}\sum_{k}(\Delta_{k}a^{\dagger}_{\vec{k}}a_{\vec{k}}-\Delta_{-k}a^{\dagger}_{-\vec{k}}a_{-\vec{k}}+\Delta_{k}b^{\dagger}_{\vec{k}}b_{\vec{k}}-\Delta_{-k}b^{\dagger}_{-\vec{k}}b_{-\vec{k}}),
	\end{aligned}
	\end{equation}
with,
	\begin{equation*}
	\Delta_{k}=2D \big[-\sin(\vec{k} \cdot \vec{a}_{1})+\sin(\vec{k} \cdot \vec{a}_{2})+\sin(\vec{k} \cdot (\vec{a}_{1}-\vec{a}_{2}))\big].
	\end{equation*}
The lattice vectors and the sign conventions in the DM coupling term are given in Fig.~\ref{fig:highsymmetry}(b).

\end{widetext}
\end{document}